\documentclass[aps,pre,twocolumn,showpacs,superscriptaddress,amsmath,amssymb,floatfix]{revtex4}
\usepackage{graphicx}
\usepackage{dcolumn}
\usepackage{bm}
\begin{document}

\title{Pair correlation functions and the wavevector-dependent surface tension in a simple density functional treatment of the liquid-vapour interface}

\author{A.O.\ Parry}
\affiliation{Department of Mathematics, Imperial College London, London SW7 2BZ, UK}

\author{C.\ Rasc\'{o}n}
\affiliation{GISC, Departamento de Matem\'aticas, Universidad Carlos III de Madrid, 28911 Legan\'es, Madrid, Spain}

\author{G.\ Willis}
\affiliation{Department of Mathematics, Imperial College London, London SW7 2BZ, UK}

\author{R.\ Evans}
\affiliation{HH Wills Physics Laboratory, University of Bristol, Bristol BS8 1TL, United Kingdom }

\begin{abstract}
We study the density-density correlation function $G({\bf r},{\bf r}')$ in the interfacial region of a fluid (or Ising-like magnet) with short-ranged interactions using square gradient density functional theory. Adopting a simple double parabola approximation for the bulk free-energy density, we first show that the parallel Fourier transform $G(z,z';q)$ and  local structure factor $S(z;q)$ separate into bulk and excess contributions. We attempt to account for both contributions by deriving an interfacial Hamiltonian, characterised by a wavevector dependent surface tension $\sigma(q)$, and then reconstructing density correlations from correlations in the interface position. We show that the standard crossing criterion identification of the interface, as a surface of fixed density (or magnetization), does not explain the separation of $G(z,z';q)$ and the form of the excess contribution. We propose an alternative definition of the interface position based on the properties of correlations between points that
"float" with the surface and show that this describes the full $q$ and $z$ dependence of the excess contributions to both $G$ and $S$. However, neither the "crossing-criterion" nor the new "floating interface" definition of $\sigma(q)$ are quantities directly measurable from the total structure factor $S^{tot}(q)$ which contains additional $q$ dependence arising from the non-local relation between fluctuations in the interfacial position and local density. Since it is the total structure factor that is measured experimentally or in simulations, our results have repercussions for earlier attempts to extract and interpret $\sigma(q)$.
\end{abstract}



\maketitle

\section{Introduction}

It is well understood that the interface between coexisting fluid phases, such as liquid and gas, is subject to thermally excited capillary-wave-like fluctuations which dominate the long-wavelength decay of density correlations \cite{Evans1990}. At distances much larger than the microscopic scale set by the bulk correlation length, long wavelength undulations in the local interfacial height $\ell({\bf x})$ increase the interfacial area and are resisted by the equilibrium surface tension $\sigma$. From this Capillary-Wave Hamiltonian picture, it follows that in the absence of additional external pinning effects, such as gravity, the thermal average of the height fluctuations satisfies \cite{BLS65}
\begin{equation}
 \langle|\,\tilde\ell({\bf q})|^2\rangle\;=\;\frac{1}{\sigma q^2}
\label{intro}
\end{equation} 
where $\tilde\ell({\bf q})$ are the Fourier components of $\ell({\bf x})$, with ${\bf x}=(x,y)$, and we have set $k_B T=1$. This is consistent with more microscopic theories which predict that, in the limit of large wavelengths, the parallel Fourier transform of the density-density correlation function behaves as \cite{Wertheim76,Weeks77,Evans1979,Evans1981,Rowlinson1982}
\begin{equation}
G(z,z';q)\;\approx\;\frac{m'(z)m'(z')}{\sigma q^2};\qquad q\to 0
\label{sumrule}
\end{equation}
for $z$ and $z'$ in the interface. Here, $m(z)$ is the equilibrium density profile, and $m'(x)\equiv dm/dz$. In recent years, several attempts have been made to refine interfacial Hamiltonian theory by allowing for a wavevector dependent surface tension $\sigma(q)$ which replaces $\sigma$ in the expression (\ref{intro}) \cite{Rochin1991,Blokhuis93,Napio93,Parry94,Mecke99,Blokhuis99,Fradin2000,Mora2003,Blokhuis2008,Blokhuis2009,Parry2008a,Hofling2014}. This quantity does not have the same thermodynamic status as the equilibrium tension, and is certainly sensitive to how one defines the interface \cite{Tarazona2007,Tarazona2012}. Nevertheless, the goal has been to find an expression for $\sigma(q)$ that can be used to calculate correlations over the range of wave-vectors $q$ reaching to the atomic scale, thus avoiding the ad hoc imposition of a high momentum cut-off in the standard capillary-wave theory expression (\ref{intro}). The purpose of the present paper is to illustrate, using a very simple square 
gradient 
density functional theory (DFT): i) Some of the subtleties involved in defining an interfacial Hamiltonian which describes not just the longest wavelength correlations, given by Eq.~(\ref{sumrule}), but also those at smaller wavelengths, and ii) to make precise the connection between correlations in the interfacial position and the density-density correlation function $G(z,z';q)$. Our DFT is sufficiently simple that we can determine exactly the full wave-vector dependence of $G(z,z';q)$ by solving the Ornstein-Zernike equation, and show that $G$ separates unambiguously into "bulk" and "excess" contributions. We then ask if both the wave-vector and position dependence of $G$ can be accounted for using effective Hamiltonian theory. This requires that we first integrate out degrees of freedom in order to derive an interfacial Hamiltonian $H[\ell]$, and then  systematically reconstruct the density correlations from the height-height correlations while allowing for a non-local relationship between height 
fluctuations and order parameter fluctuations. We show that the standard "crossing criterion" identification of the interface, as a surface of fixed density, does reproduce some properties of $G$ but does not distinguish correctly bulk from interfacial contributions. We propose an alternative definition of the interface which involves density correlations between particle positions that "float" with the surface, and which corresponds to a surface of fixed density only at very long wavelengths. This many-body definition of the interface separates bulk and interfacial modes, thus allowing us to identify the wave-vector and position dependence of the excess contributions to the pair correlation function {\it and} the local structure factor $S(q;z)$. However, the corresponding $\sigma(q)$ is not directly measurable from the local or total structure factor $S^{tot}(q)$ without allowing for the additional $q$ dependence arising from the non-local relation between fluctuations in the interfacial position and 
density. We believe this last point has implications for the way in which $\sigma(q)$ has been extracted in many earlier studies.\\

\section{Square Gradient Theory for Density Correlations: Double Parabola Approximation}

A square-gradient free-energy density functional or, equivalently, a mean-field (MF) treatment of a Landau-Ginzburg-Wilson Hamiltonian is the simplest microscopic description of the interfacial region in systems with short-ranged intermolecular forces \cite{Evans1990,Evans1979,Evans1981,Rowlinson1982}. This approach does not incorporate long-ranged attractive intermolecular forces, short-ranged volume exclusion, non-classical behaviour near the bulk critical point or capillary-wave broadening of the interface. Nevertheless, it has given invaluable insights into the nature of the interfacial region and is sufficiently simple to allow for an analytic determination of the wave-vector dependence of all quantities of interest. The free-energy functional is given by
\begin{equation}
H_{LGW}[m]=\int\!\!d{\bf r}\; \left\{\,\frac{f}{2} (\nabla m)^ 2 + \Delta \phi(m) \right\}
\label{LGW}
\end{equation}
where, adopting a magnetic language, $m({\bf r})$ is a magnetization-like order parameter. The coefficient $f$ is proportional to the second moment of the bulk direct correlation function [5] and, in its simplest approximation, is regarded as a constant \cite{Rowlinson1982}. Throughout the paper, we will use units in which $f=1$. Below a bulk critical temperature $T_c$, a suitable double-well 'potential' $\phi(m)$, which for simplicity we assume has an Ising symmetry, models the coexistence of bulk phases with order parameters $\pm m_0$. The shifted potential $\Delta\phi(m)\equiv\phi(m)-\phi(m_0)$ subtracts the bulk phase contribution from the free energy density, and the curvature $\Delta\phi '' (m_0)=\kappa^ 2$ identifies the inverse bulk correlation length $\kappa=1/\xi_b$ since, for an isotropic bulk fluid treated in mean-field, the Fourier transform of the order parameter correlation function, or bulk structure factor, arising from Eq.~(\ref{LGW}) is $S^b(Q)=(Q^2+\kappa^2)^{-1}$. We suppose that a 
planar interface of macroscopic area $A$ separates the bulk phases near the $z=0$ plane. At mean-field level, the equilibrium order-parameter profile $m(z)$ is obtained from minimizing the functional (\ref{LGW}) and satisfies the Euler-Lagrange 
equation
\begin{equation}
\frac{ d^2m}{d z^2} = \Delta \phi' (m)
\label{EL}
\end{equation}
with boundary conditions $m(\pm\infty)=\pm m_0$. This equation has a first integral, which allows us to identify the surface tension $\sigma=H_{LGW}[m(z)]/ A$ via the celebrated van der Waals formula \cite{Evans1990,Evans1979,Rowlinson1982}
\begin{equation}
\sigma\,=\int_{-\infty}^{\infty}\!\!\!dz\;m'(z) ^2\;=\;\int_{-m_0}^{m_0}\!\!\!dm\;\sqrt{2\Delta\phi(m)}
\label{tension}
\end{equation}
The mean-field result for the order-parameter correlation function
\begin{equation}
 G({\bf r},{\bf r}')\;=\;\langle m({\bf r})m({\bf r}')\rangle-m(z)m(z'),
\end{equation}
with $\langle m({\bf r})\rangle=m(z)$, follows from the solution of the Ornstein-Zernike (OZ) equation and, for the functional (\ref{LGW}), reduces to the differential equation \cite{Evans1981}
\begin{equation}
\left(-\partial^2_{z}+q^ 2 +\Delta\phi''(m(z))\right)G(z,z';q)=\delta(z-z')
\label{OZ}
\end{equation}	
where we have taken a parallel Fourier transform with respect to ${\bf x}$.  Infinitely far from the interfacial region, the correlation function must tend to the bulk result
\begin{equation}
G^b(|z-z'|;q)\;=\;\frac{1}{2\kappa_q}\,e^{-\kappa_q |z-z'|}
\label{Gbulk}
\end{equation} 
where we have abbreviated $\kappa_q\equiv\sqrt{\kappa^2+q^2}$. One useful approach to solving the OZ equation, which we will need later, is to write the solution as a spectral expansion 
\begin{equation}
G(z,z';q)\;=\;\sum_n \frac{\,\Psi_n^*(z)\Psi_n(z')}{E_n+q^2}
\label{spectral}
\end{equation}
where the normalised eigenfunctions satisfy
\begin{equation}
\left(-\partial^2_{z}+\Delta\phi''(m(z))\,\right)\,\Psi_n(z)\,=\,E_n \Psi_n(z)
\end{equation}
Inspection of the profile equation (\ref{EL}) identifies a ground state $E_0=0$ with eigenfunction $\Psi_0(z)=m'(z)/\sqrt{\sigma}$, in addition to higher energy (including scattering) states \cite{Evans1981}. Thus, for $q\ll\kappa$ the ground state dominates and $G$ behaves according to (\ref{sumrule}). We wish to go beyond this and understand the full wave-vector dependence of the pair correlation function in the interfacial region. The simplest model for the bulk free-energy which allows us to do this is the Double Parabola (DP) approximation
\begin{equation}\label{DP}
\Delta\phi(m)\;=\;\frac{\;\kappa^2}{2}\,\big(|m|-m_0\big)^2
\end{equation}
which has been used successfully in the theory of short-ranged wetting \cite{Parry2004,Parry2006a,Parry2007,Parry2008,Parry2008a}. For the DP potential, the equilibrium profile and surface tension are given by
\begin{equation}
m(z)\,=\,\textup{sign}(z)\;m_0\big(1- e^{-\kappa |z|}\big),\hspace{.75cm}\sigma\,=\,\kappa\,m_0^2
\end{equation}
which share the same qualitative features of the usual "$m^4"$ theory with the $\,\tanh(\kappa z/2)$ profile \cite{Rowlinson1982}. Within the DP approximation, the OZ equation (\ref{OZ}) can be written as
\begin{equation}
\big(-\partial^2_{z}+q^ 2 +\kappa^2 +c\,\delta(z)\big)\,G(z,z';q)\;=\;\delta(z-z')
\label{OZ2}
\end{equation}
where $c=-2\kappa$ is the coefficient of the delta function arising from the cusp in the DP potential. Rather than substitute for $c$ and solve the equation, we take an additional Fourier transform of the OZ equation (\ref{OZ2}) w.r.t.~$z$ and rearrange to obtain
\begin{equation}\label{14}
\tilde G(q_\perp,z';q)\,=\,\frac{e^{iq_\perp z'}}{q_\perp^2+q^ 2 +\kappa^2}\,-\,c\,\frac{G(0,z';q)}{\,q_\perp^2+q^ 2 +\kappa^2}
\end{equation}
where $\tilde G(q_\perp,z';q)$ is the new Fourier transform of $G(z,z';q)$. Note that it is only the presence of the constant $c$ that distinguishes (\ref{14}) from the bulk result. To see this, note that the Fourier transform w.r.t.~$z$ in (\ref{Gbulk}) yields the first term in (\ref{14}), and recall that the usual full 3D Fourier transform of the isotropic bulk correlation function is $(Q^2 +\kappa^2)^{-1}$, which depends on the total wave-number $Q=\sqrt{q_\perp^2+q^2}$. Taking the inverse Fourier transform w.r.t.~$q_\perp$ in (\ref{14}) gives
\begin{equation}
G(z,z';q)\,=\,G^b (|z-z'|;q)\,-\,c\,G(0,z';q)G^b(|z|;q)
\label{Geq}
\end{equation}
To determine $c$, we set $z=0$ and rearrange so that
\begin{equation}\label{16}
G(0,z';q)\,=\,\frac{G^b (|z'|;q)}{\,1+c\,G^b(0;q)}
\end{equation}
As we know there must be a Goldstone mode in the limit $q\to 0$, the denominator of (\ref{16}) must vanish at $q=0$. This identifies $c$ uniquely as
\begin{equation}
c\;=\;-\,\frac{1}{G^b(0;0)}
\end{equation}
which, using (\ref{Gbulk}), recovers $c=-2\kappa$, as quoted above. Re-substitution into (\ref{Geq}) then gives 
\begin{equation}
G(z,z';q)\,=\,G^b (|z-z'|;q)\,+\,\frac{G^b(|z|;q) G^b(|z'|;q)}{G^b(0;0)-G^b(0;q)}
\label{G1}
\end{equation}
Thus, $G$ clearly separates into a background contribution, equal to the bulk correlation function $G^b (|z-z'|;q)$, and an interfacial or {\it excess} contribution; i.e.~$G=G^b+G^{ex}$, with
\begin{equation}
G^{ex}(z,z';q)\;=\;\frac{G^b(|z|;q) G^b(|z'|;q)}{G^b(0;0)-G^b(0;q)}
\label{Gex}
\end{equation}
This function contains a Goldstone $q^{-2}$ singularity as $q\to 0$, and depends only on the distance of each particle to the interface. Explicitly, the excess contribution is given by
\begin{equation} 
G^{ex}(z,z';q)\;=\;\frac{\kappa(\kappa+\kappa_q)}{2\kappa_q q^2}\;e^{-\kappa_q (|z|+|z'|)}
\label{Gint}
\end{equation}
If one ignores the $q$-dependence in $\kappa_q$, and recalls that $m'(z)=\kappa m_0\,e^{-\kappa |z|}$, (\ref{Gint}) recovers the anticipated asymptotic small-$q$ behaviour given by (\ref{sumrule}). However, the consequent spatial decay is controlled in general by the wavevector dependent $\kappa_q$, rather than $\kappa$, a fact related to non-local interfacial fluctuation effects \cite{Parry2004,Parry2008,Parry2006a,Parry2008a,Fernandez2013}. When both particles are exactly at the interface ($z=z'=0$), the correlation function takes its maximum value, and the bulk and excess contributions add up to give the full correlation function
\begin{equation}
G(0,0;q)\;=\;\frac{\kappa_q+\kappa}{2q^2}\;.
\label{G00DP}
\end{equation}
The fact that $G$ separates into bulk and excess contributions means that the local structure factor, defined as
\begin{equation}
S(z;q)\;=\,\int_{-\infty}^{\infty}\!\!\!dz'\;G(z,z';q)
\end{equation}
also separates into distinct bulk and interfacial contributions. Thus, we write
\begin{equation}
S(z;q)\;=S^b(q)+S^{ex}(z;q)
\label{SDP}
\end{equation}
where $S^b(q)=\int_{-\infty}^{\infty}\!dz\; G^b(|z|;q))$ is the bulk structure factor which, for our simple DFT, is given by $S^b(q)=1/\kappa_q^2$. Of course, in a more realistic description of a fluid, $S^b(q)$ will be different in the two coexisting phases; both will exhibit a peak at $q\approx 2\pi/d$, where $d$ is an atomic diameter, arising from short-ranged (packing) effects. The all important excess contribution, containing the Goldstone mode contribution, is 
\begin{equation}
 S^{ex}(z,q)\;=\;\frac{S^b(q)G^b(|z|;q)}{G^b(0;0)-G^b(0;q)}
\label{Sint}
\end{equation}
which is given explicitly by 
\begin{equation} 
S^{ex}(z;q)\;=\;\frac{\kappa(\kappa+\kappa_q)}{\kappa_q^2\, q^2}\;e^{-\kappa_q |z|}
\label{Sint2}
\end{equation}
If one ignores the $q$ dependence of $\kappa_q$, this result is equivalent to $S^{ex}(z;q)\approx 2m_0 m'(z)/\sigma q^2$, which follows from integrating the capillary-wave result (\ref{sumrule}) or, equivalently, from the leading order contribution to the spectral expansion (\ref{spectral}). \\
Integration of $S(z;q)$, over a macroscopic range $[-L,L]$, defines the total structure factor which is given by
\begin{equation} 
S^{tot}(q)\;\equiv 2L S^b(q)+\;\frac{2\kappa(\kappa+\kappa_q)}{\kappa_q^3\,q^2}
\label{StotDP}
\end{equation}
where the second term is clearly the excess contribution, containing the Goldstone mode contribution.\\

At this stage, we might follow the standard approach outlined in the Introduction and wrap up the entire $q$ dependence of the excess piece of (\ref{StotDP}) in an effective $q$ dependent surface tension that generalizes the integral of (\ref{sumrule}); i.e.~we might choose to \textit{define} $\sigma_\textup{eff}(q)$ via
\begin{equation}
S^{tot}(q)\;=2L S^b(q)+\;\frac{4\,m_0^2}{\sigma_\textup{eff}(q)\,q^2}
\label{81}      
\end{equation}
Comparison with (\ref{StotDP}) identifies
\begin{equation}
\sigma_\textup{eff}(q)\;=\;\sigma\;\frac{2(1+q^2\xi_b^2)^2}{1+q^2\xi_b^2+\sqrt{1+q^2\xi_b^2}}
\label{82}
\end{equation}
for our DP model. As shown in Fig.~1, the function $\sigma_\textup{eff}(q)$ \textit{increases} with $q$ and, for $q\ll\kappa$, we find
\begin{equation}\label{X}
      \sigma_\textup{eff}(q)\;=\;\sigma\;\left(1+\frac{5}{4}q^2\xi_b^2+\dots\right)
\end{equation}
In the standard approach, one measures the total structure factor $S^{tot}(q)$, subtracts a suitable bulk piece which, for a fluid, would be a weighted average of the liquid and vapour bulk structure factors, and from the remaining excess piece extracts $\sigma_\textup{eff}(q)$ in the manner of (\ref{81}). Of course, for a fluid, $2m_0$ is is replaced by the difference in coexisting densities $(\rho_l-\rho_v)$. For a recent example of this methodology, see \cite{Hofling2014}, which describes results of large scale molecular dynamics simulations for the Lennard-Jones model.\\

We conclude this section by remarking that the present MF treatment of the interfacial region, of course, does not take into account the broadening of the interface region associated with capillary-wave-like fluctuations. As is well known, the MF interfacial profile $m(z)$ remains sharp ($m'(z)\ne 0$) even in the absence of an external field, while more correctly the interfacial width must diverge as $\xi_\perp \approx \sqrt{\ln A}$ in three dimensions as the interfacial area $A$ diverges \cite{Evans1990,BLS65,Wertheim76,Weeks77}. This incorrect MF treatment of interfacial wandering is manifest in the behaviour of the MF pair correlation function. While the MF expression (\ref{Gint}) for $G(z,z';q)$ is consistent with the exact sum-rule requirement (\ref{sumrule}), when we Fourier invert w.r.t.~$q$, the MF expression to determine the value of the pair correlation function $G({\bf r},{\bf r}')$ for two near-by molecules located in the interfacial region, we find that in three dimensions the result diverges as 
$\ln A$, which is unphysical. Beyond MF, this divergence is cancelled because the term  $m'(z)\sim \xi_\perp^{-1}$ vanishes as $1/\sqrt{\ln A}$. That is, the vanishing of the numerator in (\ref{sumrule}), due to capillary-wave broadening of the interface, ensures that the value of the pair correlation function between nearby molecules remains finite \cite{ParryPhD,EvansDFT1992}. Thus, without invoking any effective interfacial Hamiltonian description of the interface, one can see why the Goldstone mode singularity in $G(z,z';q)$ must lead to interfacial broadening.\\

In the remainder of the paper, we try to account for the wave-vector and position dependence of $G$ and $S$ revealed here, in particular the separation into bulk and excess contributions, using effective Hamiltonian theory. 

\section{Interfacial Hamiltonian Theory}

Any derivation of an interfacial Hamiltonian requires a suitable criterion for defining the interface position. For example, one may adopt the intuitively straightforward  \textit{crossing criterion}, which identifies $\ell({\bf x})$ as a surface of fixed isomagnetization $m({\bf x},z\!=\!\ell({\bf x}))=m^x$ [24,25]. This is a {\it{local}} definition of the interface, in the sense that $\ell({\bf x})$ depends only on the properties of the magnetization at a given point $({\bf x},\ell({\bf x}))$.  Assuming that we have a recipe for identifying the interface (not necessarily the crossing criterion), the effective Hamiltonian $H[\ell]$ is defined as the constrained minimization of (\ref{LGW}) \cite{Fisher1991,Jin1993}
\begin{equation}
H[\ell]=H_{LGW}[m_\Xi]
\label{Hdef}
\end{equation} 
where $m_\Xi({\bf r})$ is the order parameter configuration that minimizes the LGW Hamiltonian subject to the interfacial constraint (given by $\ell({\bf x})$) and bulk boundary conditions. The constrained profile satisfies the Euler-Lagrange equation 
\begin{equation}
\nabla^ 2 m_\Xi\,=\,\Delta\phi'(m_\Xi)
\label{PDE}
\end{equation}
and is a function of position, and a functional of the interfacial shape $\ell({\bf x})$. In writing (\ref{PDE}), we have assumed there is no Lagrange multiplier arising from the constraint. This is the case for the crossing criterion and other sufficiently "local" definitions of the interface position which do not involve moments of the magnetization profile over the system volume. Since we are only interested in mean-field correlations, which arise from small fluctuations, we write this constrained profile as a perturbation about the planar equilibrium profile 
\begin{equation}
m_\Xi({\bf r})= m(z)+ \delta m_\Xi(z;{\bf x})
\label{pert}
\end{equation}
The fluctuation $\delta m_\Xi(z;{\bf x})$ satisfies the linearised equation
\begin{equation}\label{30}
\nabla^ 2 \delta m_\Xi(z;{\bf x})\,=\,\Delta\phi''\big(m(z)\big)\,\delta m_\Xi(z;{\bf x})
\end{equation}
with boundary conditions that constrain $m_\Xi$ to the given profile $\ell({\bf x})$, and asymptotically $\delta m_\Xi \to 0$ as $z \to \pm\infty$. Following Parry and Boulter \cite{Parry94}, we solve (\ref{30}) using the convolution
\begin{equation}
\delta m_\Xi(z;{\bf x})\;=\;\int\!d{\bf x}'\;\Lambda (z;|{\bf x}-{\bf x}'|)\,\ell({\bf x}')
\label{Lambdadef}
\end{equation}
which, in Fourier space, reads
\begin{equation}
\delta\widetilde{m}_\Xi(z;{\bf q})\;=\;\tilde \Lambda(z;q)\,\tilde\ell({\bf q})
\label{conv}
\end{equation}
Thus, the function $\tilde \Lambda(z;q)$ satisfies 
\begin{equation}
\big(-\partial^2_{z}+q^ 2 +\Delta\phi''(m(z))\big)\,\tilde\Lambda(z;q)\,=\,0
\label{LambdaODE}
\end{equation}
which is essentially a rewrite of the OZ equation (\ref{OZ}). Equation (\ref{LambdaODE}) has local boundary conditions arising from the definition of $\ell(\bf x)$ (see later), and also bulk conditions $\tilde\Lambda(z;q)\to 0$ as $z\to\pm\infty$. Already from the Euler-Lagrange equation (\ref{EL}) for the planar profile, we see that the zeroth moment satisfies
\begin{equation}
\tilde\Lambda(z;0)=-m'(z)
\label{lambda0}
\end{equation}
consistent with a pure translation of the planar equilibrium profile $m(z)\to m(z-\ell)$. Substituting the convolution (\ref{Lambdadef}) into (\ref{Hdef}) leads to the extended capillary-wave Hamiltonian 
\begin{equation}
H[\ell]\;=\;\frac{1}{2}\,\sum_{{\bf q}}\, \sigma(q)\,|\tilde \ell({\bf q})|^2 
\label{ECW}
\end{equation}
where the sum extends over all wavevectors up to the cut-off provided by the LGW model, and the wave-vector-dependent surface tension is identified as
\begin{equation}
\sigma(q)\;=\,-\int\! dz\; m'(z)\,\tilde\Lambda(z;q)\,.
\label{sigmaq}
\end{equation}
It follows from (\ref{lambda0}) that this result reduces to the equilibrium surface tension, $\sigma(q\to 0)=\sigma$, as required on physical grounds. Since $H[\ell]$ is quadratic in $\tilde \ell({\bf q})$, the equilibrium height-height correlations are given by
\begin{equation}
 \langle|\,\tilde\ell({\bf q})|^2\rangle_\ell\;=\;\frac{1}{\sigma(q)\, q^2}
\label{hhsq}
\end{equation}
which is the sought-after generalization of (\ref{intro}). The subscript on the equilibrium bracket is a reminder that this average is now defined w.r.t.~the interfacial Hamiltonian (\ref{ECW}) rather than the microscopic Hamiltonian (\ref{LGW}).\\

The above theory also allows us to solve the inverse problem of obtaining a {\it reconstructed} order-parameter correlation function
\begin{equation}\label{Grecondef}
\mathcal{G}({\bf r},{\bf r}')\,=\,\langle \delta m_\Xi({\bf r})\delta m_\Xi({\bf r}')\rangle_\ell-\langle \delta m_\Xi({\bf r})\rangle_\ell\langle \delta m_\Xi({\bf r}')\rangle_\ell
\end{equation}
from the interfacial description given by the Hamiltonian (\ref{ECW}). Note that, because the constrained profile $m_\Xi$ necessarily tends to the appropriate bulk magnetization as $z\to\pm\infty$, the reconstruction of the correlation function is necessarily incomplete and, indeed, $\mathcal{G}({\bf r},{\bf r}')$ will not recover the bulk correlation function $G^b(|{\bf r}-{\bf r}'|)$ as one moves away from the interface. Rather, it must vanish. We shall return to this shortly.\\

In general, the reconstruction of correlations via (\ref{Grecondef}) is a non-trivial task because $m_\Xi({\bf r})$ is a functional of the interfacial shape. However, at mean-field level, the convolution (\ref{Lambdadef}) provides the link between order-parameter and interfacial fluctuations, and we can write
\begin{equation}
\mathcal{G}(z,z';q)\;=\;\frac{\tilde \Lambda(z;q)\tilde\Lambda(z';q)}{\sigma(q)\, q^2}
\label{Grecon}
\end{equation}
where a parallel Fourier transform w.r.t.~${\bf x}$ has been performed. Expression (\ref{Grecon}) demonstrates that the $q$-dependence in the correlation function arises from both $\sigma(q)$ {\it and} the function $\tilde\Lambda(z;q)$ that describes the non-local relation between interfacial and order parameter fluctuations. The reconstructed pair correlation function depends only on the distance of each particle to the interface and, therefore, provides some measure of the "interfacial", preferably what we have identified as the excess, contribution to the pair correlation function $G$. This is to be expected, of course. Integrating out degrees of freedom, equivalent to performing a constrained minimization, must lead to a loss of information, so one cannot expect that $\mathcal{G}(z,z';q)$ recovers all the information that is in the full pair correlation function. However, at the very least, it does capture the correct long-wavelength behaviour occurring for $q\ll\kappa$,
where we can approximate $\tilde\Lambda(z;q)\approx -m'(z)$, and 
hence $\sigma(q)\approx \sigma$, so that
\begin{equation}
\mathcal{G}(z,z';q)\,\approx\; \frac{m'(z)m'(z')}{\sigma\, q^2}
\label{Greconlim}
\end{equation}
for \textit{any} choice of the double-well potential $\Delta\phi(m)$. Thus, the effective Hamiltonian theory recovers the correct long wavelength behaviour of the correlation function near the interface. However, $\mathcal{G}(z,z';q)$ is not equivalent to the full $G(z,z';q)$ and, as remarked above, does not reproduce the bulk correlation function (\ref{Gbulk}) when the particles are far from the interface. To proceed further, we now consider the crossing criterion in more detail.

\section{Correlation function structure using a crossing criterion}

Within the crossing criterion (cc), the interface is defined as a surface of constant, fixed order-parameter $m^x=0$. Thus, we have
\begin{equation}
m_\Xi({\bf r}_\ell)\,=\,0
\label{CC}
\end{equation}
for all points ${\bf r}_\ell=({\bf x},\ell({\bf x}))$ along the interface. If we start from the equilibrium planar configuration at $z=0$ and consider a small height fluctuation, the magnetization at the new interface position, $m(\ell({\bf x}))+\delta m_\Xi ({\bf x},\ell({\bf x}))$, remains zero. Expanding to first-order in $\ell({\bf x})$ gives 
\begin{equation}
m'(0)\,\ell({\bf x})\,+\,\int\! d{\bf x}'\;\Lambda_{cc}(0;|{\bf x}-{\bf x}'|)\,\ell({\bf x}')\;=\;0,
\label{CCexp}
\end{equation}
where we use the subscript $cc$ to specify we have imposed the cc, and yields the boundary condition \cite{Parry94}
\begin{equation}
\tilde \Lambda_{cc} (0;q)=-m'(0)\,.
\label{lambdabc}
\end{equation}
Thus, from (\ref{Grecon}) the expression for the reconstructed correlation function exactly at the interface simplifies to
\begin{equation}
\mathcal{G}_{cc}(0,0;q)\,=\,\frac{\,m'(0)^2}{\sigma_{cc}(q)\, q^2}
\label{Grecon00}
\end{equation}
while the function $\tilde\Lambda_{cc}(z;q)$ can be written in terms of the equilibrium pair correlation function as
\begin{equation}
\tilde\Lambda_{cc}(z,q)\,=\,-\,m'(0)\,\frac{G(0,z;q)}{\,G(0,0;q)},
\label{Lambdag-}
\end{equation}
which follows from comparing (\ref{OZ}) and (\ref{LambdaODE}). We now substitute into the expression (\ref{sigmaq}) for $\sigma(q)$ and use the spectral expansion (\ref{spectral}) for $G(0,z;q)$. Noting that $m'(z)=\sqrt{\sigma}\;\Psi_0(z)$, orthonormality cancels all but the ground-state contribution, implying that
\begin{equation}
\mathcal{G}_{cc}(0,0;q)=G(0,0;q)
\label{Grecon=G}
\end{equation}
Thus, within the crossing criterion, the reconstructed correlation function $\mathcal{G}_{cc}$ reproduces exactly the solution to the OZ equation when both particles are at the interface - a result \textit{not restricted} to the DP potential. By virtue of (\ref{Lambdag-}), this also means that
\begin{equation}
\mathcal{G}_{cc}(0,z;q)=G(0,z;q)
\label{Grecon=G2}
\end{equation}
and also that $\mathcal{G}_{cc}(z,z';q)=G(z,z';q)$ if $zz'<0$, i.e.~when the particles are on {\it opposite} sides of the interface. However, when both particles are at the {\it same} side of the interface ($zz'>0$), the functions $\mathcal{G}_{cc}$ and $G$ differ. If we define a reconstructed structure factor $\mathcal{S}(z;q)$ as the integral of (\ref{Grecon}) over all $z'$, it follows from (\ref{Grecon=G2}) that the reconstructed function, based on the cc, is equal to the full structure factor at the interface, where $z=0$:
\begin{equation}
\mathcal{S}_{cc}(0;q)=S(0;q)
\label{Srecon=S}
\end{equation}
Although the three results above appear to demonstrate a certain degree of success of the cc definition of the interface, since this reproduces exactly some properties of the full correlation function $G$, they conceal a highly unsatisfactory and physically contradictory element. To see this, recall that the reconstructed function $\mathcal{G}(z,z';q)$ vanishes as $|z|,|z'| \to \infty$ and, therefore, cannot contain a bulk contribution far from the interface. This is the reason why $\mathcal{G}(z,z';q)$ differs from the full $G(z,z';q)$ when the particles are on the same side of the interface. Thus, as stressed earlier, our reconstructed function $\mathcal{G}(z,z';q)$ can only be interpreted, strictly speaking, as a measure of the {\it excess} contribution to $G$. With this in mind, we see that (\ref{Grecon=G})-(\ref{Srecon=S}) actually obscure the physics, since we know from the DP results (\ref{G1}) and (\ref{SDP}) that $G(0,0;q)$ and $S(0;q)$ contain bulk contributions. A physically consistent 
interpretation is therefore only achieved if the reconstructed correlation function and structure factor are their corresponding excess (interfacial) contributions.  The upshot is two fold: Firstly, the prediction of the cc for the reconstructed structure factor $\mathcal{S}(z;q)$ is inaccurate away from the interface. Secondly and more importantly, we believe that the cc definition of $\sigma(q)$ is, for these purposes and certainly for considering fluctuation effects beyond mean-field, artificial.\\

Let us check these predictions based on the cc against the explicit results derived earlier for the DP potential. The comparison illustrates the highlighted problems. The OZ-like equation (\ref{LambdaODE}) for the the function $\tilde\Lambda(z;q)$ reduces to
\begin{equation}
\big(-\partial^2_{z}+q^ 2 +\kappa^2\big)\,\tilde\Lambda(z;q)\,=\,0
\label{LambdaODEDP}
\end{equation}
so that
\begin{equation}
\tilde\Lambda_{cc}(z;q)\,=\,-m'(0)\,e^{-\kappa_q|z|}
\end{equation}
with $m'(0)=\kappa\, m_0$. Substituting into (\ref{sigmaq}) determines the wave-vector dependent surface tension as
\begin{equation}
\sigma_{cc}(q)\,=\,\frac{2\sigma}{\,1+\sqrt{1+q^2\xi_b^2}},
\label{sigmaCC}
\end{equation}
which, in contrast to $\sigma_\textup{eff}(q)$ in (\ref{82}), decreases with increasing $q$ (See Fig.~1). Substituting $\sigma_{cc}(q)$ into (\ref{Grecon00}) then gives 
\begin{equation}
\mathcal{G}_{cc}(0,0;q)=\frac{\kappa_q+\kappa}{2q^2}
\end{equation}
which, in accordance with (\ref{Grecon=G}), is the same as the result (\ref{G00DP}) for the full pair correlation function found from the original OZ equation. For arbitrary $z$ and $z'$, the reconstructed correlation function takes the form 
\begin{equation}
\mathcal{G}_{cc}(z,z';q)\;=\;\frac{\kappa_q+\kappa}{2q^2}\,e^{-\kappa_q (|z|+|z'|)}
\end{equation}
This is \textit{not} the excess contribution (\ref{Gint}). Rather 
\begin{equation}
\mathcal{G}_{cc}(z,z';q)\;=\;\frac{1}{2\kappa_q}\;e^{-\kappa_q (|z|+|z'|)}\,+\,G^{ex}(z,z';q)
\label{Greconcc}
\end{equation}
Thus, if $zz'=0$, or if the particles are on opposite sides of the interface, we recover $\mathcal{G}_{cc}=G$, since the first term in (\ref{Greconcc}) then becomes the bulk pair correlation function. However, in general, the reconstructed correlation function is neither the full $G$ nor the excess contribution $G^{ex}$. Similar statements apply to the reconstructed structure factor which, according to the crossing-criterion is given by integrating (\ref{Greconcc}) over $z'$:
\begin{equation}\label{55}
\mathcal{S}_{cc}(z;q)\;=\;\left(\frac{1}{\kappa_q^2}+\;\frac{\kappa(\kappa+\kappa_q)}{\kappa_q^2 q^2}\right)\;e^{-\kappa_q|z|}
\end{equation}
which agrees with the result for $S(z;q)$ only at the interface, $z=0$; see (\ref{SDP}) and (\ref{Sint2}). However in general, the reconstructed structure factor is neither the excess contribution $S^{ex}$ nor the full structure factor $S$. This error then feeds into the total structure factor obtained by integrating ${S}_{cc}(z;q)$ which, within the cc is 
\begin{equation}
\mathcal{S}_{cc}^{tot}(q)\;=\;\frac{2}{\kappa_q^3}+\;\frac{2\kappa(\kappa+\kappa_q)}{\kappa_q^3 q^2}
\label{stotbad}
\end{equation}
which should be compared to the correct result for $S^{tot}(q)$ given by (\ref{StotDP}). Clearly, (\ref{stotbad}) does not yield the correct excess contribution to $S^{tot}(q)$ since it contains an unwanted additional contribution $2/\kappa_q^3$ which is dominant at high wavevectors $q>\kappa$. Thus (\ref{stotbad}) fails completely to describe the $q$ dependence over the entire wavevector regime.\\

\section{Beyond the crossing criteria: A Floating correlation function definition of the interface}

The above analysis highlights the inadequacy of using a crossing criterion as the definition of the interface, since there is no separation of interfacial and bulk contributions to the pair correlation function. This is not an issue if we are only interested in very long wavelength behaviour occurring for $q\ll\kappa$.  However, it implies that there is probably no physical significance to the wave-vector dependent surface tension (\ref{sigmaCC}). At the very least, it is not a useful means of understanding the wavevector dependence of the total structure factor. Clearly, it would be more meaningful to have a definition of the interface in which the reconstructed pair correlation function is just the excess contribution $G^{ex}$ (\ref{Gint}), or one for which the reconstructed structure factor is its excess contribution $S^{ex}$ (\ref{Sint2}).\\
To do this, we focus on correlations along the interface itself. Within the crossing criterion, the magnetization at any point on the interface is fixed to $m^x=0$. Therefore, the correlation function between any pair of points ${\bf r}_\ell=({\bf x},\ell({\bf x}))$  and ${\bf r}'_\ell=({\bf x}',\ell({\bf x}'))$ which "float" with the interface is trivially zero: $\langle m_\Xi({\bf r}_\ell)m_\Xi({\bf r}'_\ell)\rangle_\ell =0$. Instead, we seek a definition of the interface which accounts for the background structure in this correlation function by allowing the value of the magnetization along the surface to fluctuate. For a planar interface ($\ell({\bf x})=0$), we assume that the magnetization along it takes the value zero, say. When the interface fluctuates, we impose that
\begin{equation}
\langle m_\Xi({\bf r}_\ell)\rangle_\ell=0
\end{equation}
and
\begin{equation}
\langle m_\Xi({\bf r}_\ell)m_\Xi({\bf r}'_\ell)\rangle_\ell=g(|{\bf x}-{\bf x}'|)
\label{Gxi}
\end{equation}
Thus, instead of it being a surface of fixed magnetization, the interface is considered to be a surface along which the {\it average} magnetization is zero, {\it but} which has a specific correlation structure given by the function $g(|{\bf x}-{\bf x}'|)$ to be determined. This does not introduce a Lagrange multiplier into the Euler-Lagrange equation (\ref{PDE}) since no integral or moment of the magnetization profile over the system volume is involved. The result (\ref{sigmaq}) for $\sigma(q)$ and (\ref{hhsq}) for the height-height correlations are therefore still valid although the presence of a non zero $g$ changes the boundary conditions on the function $\Lambda(z;q)$ and, hence, alters the reconstruction of pair correlations. To see this, note that for small deviations from the plane, the magnetization at a point on the interface must have Fourier amplitudes $\tilde m_\Xi({\bf q})=m'(0)\tilde \ell({\bf q})+\tilde\Lambda(0;q)\tilde\ell({\bf q})$. This identifies the Fourier transform of the floating 
correlation function as
\begin{equation}
\widetilde{g}(q)\;=\;(m'(0)+\tilde\Lambda(0,q))^2\,\langle|\tilde\ell({\bf q})^2|\rangle_\ell
\end{equation}
and thus
\begin{equation}
\widetilde{g}(q)\;=\;\frac{(m'(0)+\tilde\Lambda(0,q))^2}{\sigma(q)\,q^2}
\label{Gxi2}
\end{equation}
Next, we suppose, by analogy with (\ref{lambdabc}), that the new boundary condition on the function $\Lambda(z;q)$ is
\begin{equation}
\tilde\Lambda(0;q)\;=\,-\,b(q)\,m'(0)
\label{Lambdabcnew}
\end{equation}
where $b(q)$ is to be determined from the properties of $g(q)$. Note that we must have $b(0)=1$, so that any new definition of the interface position does not alter the identification $\tilde\Lambda(z,0)=-m'(z)$ and hence $\sigma(0)=\sigma$ -- see (\ref{lambda0}) and (\ref{sigmaq}). The factor $b(q)$ simply multiplies our previous results $\tilde\Lambda_{cc}(z,q)$ and $\sigma_{cc}(q)$ derived using the crossing criterion. Thus, in an obvious notation, the new expressions are
\begin{equation}
\sigma(q)=b(q)\,\sigma_{cc}(q),\hspace{1cm} \tilde\Lambda(z,q)=b(q)\,\tilde\Lambda_{cc}(z,q)
\label{brescale}
\end{equation}
Substitution into (\ref{Gxi2}) gives 
\begin{equation}
\widetilde{g}(q)\;=\;\frac{(1-b(q))^2}{b(q)}\,\frac{m'(0)^2}{\sigma_{cc}(q)\,q^2}
\label{Gxi3}
\end{equation}
and hence, using (\ref{Grecon00}) and (\ref{Grecon=G}),
\begin{equation}
\widetilde{g}(q)\;=\;\frac{(1-b(q))^2}{b(q)}\;G(0,0;q)
\label{Gxi4}
\end{equation}
Recall that the crossing criterion recovers the structure of $G$ at the interface. Finally, the rescalings (\ref{brescale}) imply, through (\ref{Grecon}) and (\ref{Lambdabcnew}), that the reconstructed pair correlation function at the interface is 
\begin{equation}
\mathcal{G}(0,0;q)=b(q)\,G(0,0;q)
\label{Greconb}
\end{equation}
and, therefore, is no longer equal to the full pair correlation function $G$ at the interface. Equations (\ref{Gxi3}) and (\ref{Gxi4}) establish the link between the reconstructed correlation function $\mathcal{G}(0,0;q)$, the full correlation function $G(0,0;q)$ and the floating correlation function $g(q)$; they are valid for any choice of the function $b(q)$. From (\ref{Gxi3}), we see that setting $\widetilde{g}(q)=0$ implies $b(q)=1$ and recovers our previous results derived using the crossing criterion. However, other choices for $b(q)$ and $g(q)$ are certainly possible depending on what choice of "background" correlation function at the interface one wishes to allow for.\\

\subsection{Matching the Correlation Function}\label{methodA}

Suppose that we \textit{choose} to have a background contribution at the interface $z=z'=0$ such that the reconstructed correlation function $\mathcal{G}$ matches the exact excess contribution:
\begin{equation}
\mathcal{G}_G(0,0;q)\,=\,G^{ex}(0,0;q)\,=\,G(0,0;q)-G^b(0;q)
\end{equation}
where the subscript $G$ denotes matching $G$. The appropriate function $b$ for this choice follows from Eq.~(\ref{Greconb}):
\begin{equation}
b_G(q)=1-\frac{G^b(0;q)}{G(0,0;q)}
\label{b}
\end{equation}
which, in turn, determines the floating correlation function $\tilde g_G(q)$ via (\ref{Gxi4})
\begin{equation}
G(0,0;q)=G^b(0;q)+\frac{G^b(0;q)^2}{\widetilde{g}_G(q)}
\end{equation}
This decomposition of the pair correlation function in the interface is ideally suited for comparison with the DP model, and (\ref{G1}) identifies immediately the floating correlation function for this case as
\begin{equation}
\widetilde{g}_G(q)=G^b(0;0)-G^b(0;q)
\label{GllDP}
\end{equation}
The fact that this function vanishes for $q\to 0$ means that, for long length scales, the interface still has its intuitive interpretation as a surface of constant magnetization. However, more generally, the Fourier inverse of (\ref{GllDP}) tells us that our interface must now be regarded as a surface of anti-bulk correlation. For the DP potential, the identification (\ref{b}) is particularly simple and, using (\ref{G00DP}), we find
\begin{equation}
b_G(q)=\frac{\kappa}{\kappa_q}
\end{equation}
which clearly satisfies our requirement that $b(0)=1$. Explicitly, the new results for $\sigma(q)$ and $\tilde\Lambda(z;q)$ for our DP model potential are
\begin{equation}
\sigma_G(q)=\,\frac{2\sigma}{1+q^2\xi_b^2+\sqrt{1+q^2\xi_b^2}}
\label{sigmaqnew}
\end{equation} 
and
\begin{equation}\label{72}
\tilde\Lambda_G(z;q)\,=\,-\,\frac{m'(0)}{\sqrt{1+q^2\xi_b^2\,}}\;\,e^{-\kappa_q|z|}
\end{equation}
Substitution of (\ref{72}) into (\ref{Grecon}) shows that the new reconstructed correlation function is
\begin{equation} 
\mathcal{G}_G(z,z';q)\;=\;G^{ex}(z,z';q),
\label{G4}
\end{equation}
i.e.~we recover precisely the excess contribution (\ref{Gint}) to the pair correlation function for {\it all } positions $z,z'$.\\

\subsection{Matching the Structure Factor}

Suppose now that we choose an alternative approach such that the reconstructed structure factor $\mathcal{S}(0;q)$ at the interface matches the exact excess quantity, i.e.
\begin{equation}
\mathcal{S}_S(0;q)\,=\,S^{ex}(0;q)\,=\,S(0;q)-S^b(q)
\end{equation}
where the subscript $S$ denotes matching $S$. Under the rescalings (\ref{brescale}), the reconstructed structure factor relates to that of the cc definition via
\begin{equation}
\mathcal{S}(z;q)\;=\;b(q)\,\mathcal{S}_{cc}(z;q)
\label{Srescale}
\end{equation}
and hence, at the interface $z=0$, satisfies
\begin{equation}
\mathcal{S}(0;q)\;=b(q)\;S(0;q)
\end{equation}
since $\mathcal{S}_{cc}(0;q)=S(0;q)$ -- see (\ref{55}). Hence, the function $b(q)$ for this particular choice is
\begin{equation}
b_S(q)=1-\frac{S^b(q)}{S(0;q)}
\end{equation}
Note that although this is, in principle, a different approach from method A, a remarkable feature is that, for the DP model potential, one finds
\begin{equation}
b_S(q)=b_G(q)
\end{equation}
so that that both matching criteria yield an {\it identical} floating correlation function $g(|\bf x|)$ and wavevector dependent surface tension $\sigma(q)$. This means that, using the same factor $b(q)=\kappa/\kappa_q$, we obtain a reconstructed structure factor
\begin{equation}
\mathcal S(z;q)\;=\;\frac{\kappa(\kappa+\kappa_q)}{\kappa_q^2\, q^2}\;\,e^{-\kappa_q |z|}
\label{Sintrecon}
\end{equation}
which is precisely $S^{ex}(z;q)$, the desired excess contribution to the structure factor given by (\ref{Sint2}). Obviously, integration of (\ref{Sintrecon}) recovers the second term of $S^{tot}(q)$ given by (\ref{StotDP}) \\

In Fig.~1, we plot $\sigma_S(q)=\sigma_G(q)$, given by Eq.~(\ref{sigmaqnew}). Although this function is clearly different from $\sigma_{cc}(q)$, it does decrease with increasing $q$.

\begin{figure}[h]\label{one}
\includegraphics[width=0.9\columnwidth]{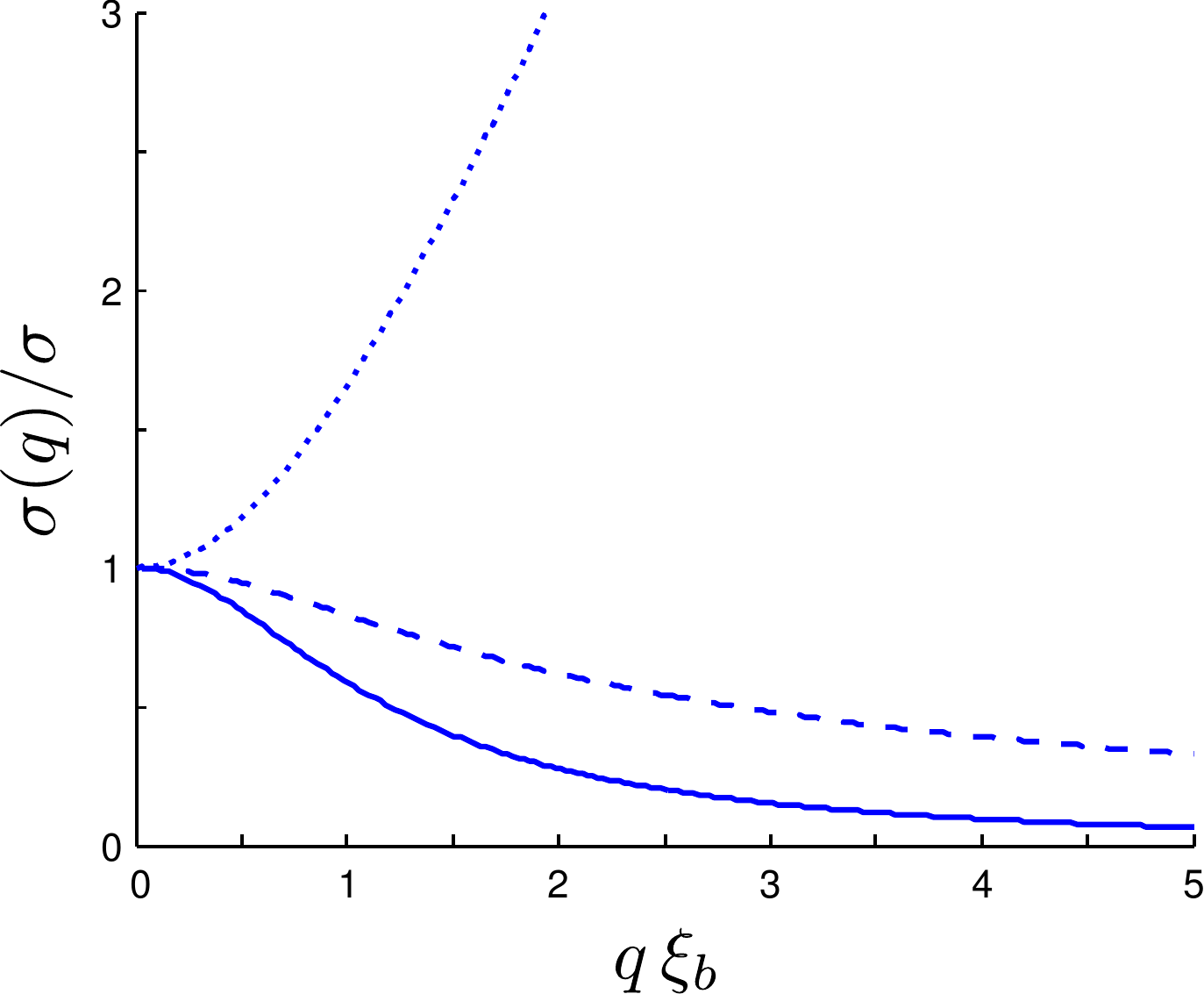}
\caption{Comparison of different wavevector dependent surface tensions $\sigma(q)$: the effective tension $\sigma_\textup{eff}(q)$ given by eq.~(\ref{82}) as defined from the excess contribution to the total structure factor (dotted line), $\sigma_{cc}(q)$ given by eq.~(\ref{sigmaCC}) as defined using the crossing criterion (dashed line), and the result from the floating point definition of the interface $\sigma_G(q)$, eq.~(\ref{sigmaqnew}), (continuous line). }
\end{figure}

\section{Discussion}

In this paper, we have discussed the structure of the pair correlation function and the identification of a wavevector dependent surface tension in a very simple square-gradient DFT treatment of a liquid-vapour interface. The DFT is far too simple to provide a quantitatively accurate microscopic description of the interfacial region. Its main drawback is the failure to describe the first peak of the bulk structure factor which requires a proper non-local treatment of short-ranged correlations \cite{Evans1990}. Nevertheless, because all quantities can be determined analytically, this approach allows us to test the physical meaning and robustness of "$\sigma(q)$" which has broader implications for theory, simulation studies and experiments. On this matter there is both good and bad news:\\

1) {\bf Good news}: Within the DP approximation (\ref{DP}), direct solution of the OZ equation shows that the pair correlation function $G(z,z;q)$, local structure factor $S(z;q)$ and total structure factor $S^{tot}(q)$ split unambiguously into bulk and excess contributions. On its own, this does not identify what $\sigma(q)$ should be but raises hope that it may be possible to derive an interfacial Hamiltonian $H[\ell]$ which explains the wavevector dependence of $G$ and $S$.\\

2) {\bf Bad news}:  The form of the excess contributions $G^{ex}$ and $S^{ex}$ cannot be accounted for using a crossing criterion (cc) definition of the interface, since the correlation function $\mathcal{G}_{cc}$ reconstructed from height-height fluctuations mixes bulk and interfacial fluctuations. Thus, while a $\sigma_{cc}(q)$ can be defined, we do not believe it has any physical meaning; i.e.~it cannot be used to compute physical observables.\\

3) {\bf Good news}: It is possible to reconstruct the correct excess contribution to $G(z,z';q)$ by introducing a new definition of the interface which allows for correlations between points that float with the interface. By allowing for such background structure one can simultaneously distinguish the excess and bulk contributions to the pair correlation function {\it and} to the structure factor and thereby obtain a more meaningful $\sigma(q)$. The new wave-vector dependent surface tension $\sigma_G(q)$ is quite different from that obtained using the crossing criterion $\sigma_{cc}(q)$, as illustrated in Fig 1. At small wave-vectors, one may compare the expansions: $\sigma_G(q)\approx\sigma(1-\frac{3}{4}q^2\xi_b^2+\cdots)$ whereas $\sigma_{cc}(q)\approx\sigma(1-\frac{1}{4}q^2\xi_b^2+\cdots)$. The difference is even more pronounced at higher wavevectors. We emphasize that both criteria yield $\sigma(q)$ that decrease with $q$.
This is in complete contrast to the standard approach which yields $\sigma_\textup{eff}(q)$ increasing with $q$ -- see (\ref{82}), (\ref{X}) and Fig.~1.\\ 

4) {\bf Very bad news}: Regardless of the definition of the interface, it is not possible, as has been commonly assumed, to extract $\sigma(q)$ from the measured total structure factor $S^{tot}(q)=\int_{-L}^{L} dz S(z,q)$. This is because the $q$ dependence arises from both the wavevector-dependent surface tension $\sigma(q)$ \textit{and} the non-local relation between interface and density fluctuations described by the function $\tilde\Lambda(z,q)$. For example, using (\ref{sigmaqnew}), the result from the floating correlation function definition, the total structure factor (\ref{StotDP}) can be expressed as 
\begin{equation}
S^{tot}(q)=2LS^b(q)+\frac{4m_0^2}{\sigma_G(q)q^2}\frac{S^b(q)^2}{S^b(0)^2}
\end{equation}
where $2L$ is the integration range (considered macroscopic) and $S^b(q)=1/(\kappa^2+q^2)$ is the bulk structure factor. Unless one knows a priori that there is a multiplicative term  $(S^b(q)/S^b(0))^2$ in the excess contribution, it is impossible to extract $\sigma_G(q)$. Note that the same problem arises if one uses the simpler crossing criterion definition of the interface in which case the multiplicative factor in the excess part is $(S^b(q)/S^b(0))^{3/2}$.\\

The last of our four summary points is perhaps the most important since it has direct implications for experimental and simulation studies that rely on the total structure factor to access the wavevector dependent tension. Of course, one may always \textit{define} a wavevector-dependent tension according to (\ref{81}). This may appear attractive since it is the total structure factor that is most easily accessible experimentally. In addition, from a theoretical perspective, such a definition does not require us to worry about the meaning of the interface position $\ell({\bf x})$ since, in principle, the structure factor can be obtained directly from solving the OZ equation; this is the route taken in (MF) DFT approaches. Thus, one may sidestep entirely the derivation of an interfacial Hamiltonian and the discussion of the meaning of $\ell({\bf x})$. However, as demonstrated above, the resulting effective tension $\sigma_\textup{eff}(q)$ is unrelated to that appearing in an interfacial Hamiltonian. Therefore, 
one is not at liberty to draw implications for the fluctuation properties of an underlying interface. We believe these cautionary remarks are not restricted to the present square-gradient DP approximation. \\

We end our article with some further discussion beginning by mentioning the inclusion of long-ranged intermolecular forces. The solution (\ref{G1}) of the OZ equation is also valid in extensions of the LGW model which include higher order gradient terms and a long-ranged intermolecular potential $w(|{\bf r}-{\bf r}'|)$; for example, the free-energy functional
\begin{equation}
\begin{array}{l}
H_{LGW}[m]=\displaystyle\int\!\!d {\bf r}\, \left\{\frac{1}{2} (\nabla m)^ 2 + \frac{\;\kappa^2}{2}\left(|m|-m_0\right)^2 \right\}\\[.4cm]
\qquad\quad+\;\displaystyle\frac{1}{2}\int\!\!\!\int\!\! d{\bf r}\,d{\bf r}' \;\,m({\bf r})\,w(|{\bf r}-{\bf r}'|)\,m({\bf r}').
\end{array}
\end{equation} 
Note that the bulk free-energy density retains the DP form. The only difference is that the bulk correlation function $G^b(z;q)$ appearing in (\ref{G1}) is more complicated and is obtained as the Fourier inverse w.r.t.~$q_{\perp}$ of 
\begin{equation}
S^b(Q)=\frac{1}{\kappa^2+Q^2+\tilde w(Q)}
\label{SbLR}
\end{equation}
where $Q=\sqrt{q^2+q_\perp^2}$ and $\tilde w(Q)$ is the 3D Fourier transform of $w(r)$. Let us examine the consequences for the structure of the correlation function in the bulk and near the interface for systems with a Lennard-Jones-like attractive tail
\begin{equation}
w(r)=-\frac{\epsilon\,a_\textup{LJ}^6}{(r^2+a_\textup{LJ}^2)^3}
\end{equation}
where $\epsilon$ is the interaction strength and $a_\textup{LJ}$ is a short-ranged cut-off. The 3D Fourier transform of this function is given explicitly by
\begin{equation}
 \tilde w(Q)= -\,\epsilon\,a_\textup{LJ}^3\,\frac{\,\pi^2}{4}\left(1+a_\textup{LJ}\,Q\right)\,e^{-a_\textup{LJ} Q}
\end{equation}
which has the small-vector expansion
\begin{equation}
\tilde w(Q)=-\,\epsilon\,a_\textup{LJ}^3\,\frac{\,\pi^2}{4}\left(1-\frac{1}{2}\,a_\textup{LJ}^2\,Q^2+\frac{1}{3}\,a_\textup{LJ}^3\, Q^3+\cdots\right)
\end{equation}
Crucially, this contains a term $\mathcal{O}(Q^3)$, absent for systems with strictly short-ranged forces, which feeds directly into the denominator of $S^b(Q)$. The consequences of this are well known and mean that, away from the bulk critical point, the pair correlation function $G^b(r)$ decays asymptotically with the same algebraic power-law as the intermolecular potential $w(r)$; i.e.~$G^{b}(r)\approx-(S^b(0))^2\omega(r)$, for $r\gg\xi_b$ \cite{Enderby1965}.
This is in sharp contrast with the exponential Ornstein-Zernike-like decay $G^b(r)= e^{-\kappa r}/4\pi r$, appropriate for systems with strictly short-ranged forces, described by (\ref{LGW}). We emphasise that these are rather general conclusions which are believed valid beyond the present mean-field analysis. The long-ranged decay of the bulk pair correlation function $G^b(|{\bf r}_1-{\bf r}_2|)\approx |{\bf r}_1-{\bf r}_2|^{-6}$ now has direct consequences for the properties of the pair correlation function in the interfacial region, which can be read immediately from (\ref{G1}). To see this, we determine the properties of the 2D 
parallel Fourier transform of the bulk pair correlation function or, equivalently, of $w(r)$. If the two particle positions have different $z$ coordinates, the zeroth moment must decay as $G^b(z;0)\sim |z|^{-4}$ when $|z|\to\infty$, which controls the spatial decay of the excess contribution to $G(z,z';q)$ for small wave-vectors. 
It follows that (\ref{Gex}) is entirely in keeping with the Capillary-Wave result (\ref{sumrule}), since it is known that the interfacial density profile decays to the bulk algebraically as $m(z)=\pm m_0+\mathcal{O}(|z|^{-3})$ for systems with dispersion interactions \cite{Rowlinson1982,Lu1985}.\\

Now, suppose that both particles have the same $z$ coordinate and consider the 2D Fourier transform of the intermolecular potential which is determined by the integral $\int_0^{2\pi}\!\!d\theta\int_0^\infty \!\!w(r)\,e^{iqr\cos\theta} r dr$. Expanding in $q$, we see that the zeroth and second moments of $w(r)$ exist but that the fourth does not because of the large distance marginal singularity $\int_0^R \!\!r^4 w(r) r d r\propto \ln R$. In turn, this means that the 2D Fourier transform of the bulk correlation function $G^b(0;q)$ has the moment expansion
\begin{equation}
G^b(0;q)=G^b_0+G^b_2\,q^2+\gamma\, q^4\ln(q\,a_\textup{LJ}\!)+\cdots
\label{GLRq}
\end{equation}
where, by definition, $G^b_0$ and $G^b_2$ are the zeroth and second parallel moment of $G^b(|{\bf{r}}_1-{\bf{r}}_2|)$ with $z_1=z_2$. These moments depend on both the short and long-ranged parts of the intermolecular interaction and are well-behaved. The non-analytic next order term in (\ref{GLRq}) arises directly from the Fourier transform of $w(r)$ with the log correction reflecting the marginal singularity of the fourth moment integral. Thus, the coefficient $\gamma$ is proportional to $\epsilon$, the strength of the intermolecular potential. \\

It follows that the denominator in (\ref{Gex}) has a small $q$ expansion proportional to $ q^2\left(1+(\gamma/G^b_2)(q^2\ln q)+...\right)$ which demonstrates the presence of: a) a Goldstone $q^{-2}$ divergence of $G^{ex}$ as $q\to 0$, and b) a non-analytic logarithmic correction arising from the long-ranged forces. This non-analyticity agrees with the predictions for the wave-vector dependent surface tension in systems with dispersion forces by Mecke and Dietrich \cite{Mecke99}. However,  we have \textit{not} had recourse to define a wave-vector dependent surface tension. Rather, we determine directly the properties of the excess part of the pair correlation function and structure factor. This suggests that the non-analytic {\it low} wave-vector correction to the surface tension is insensitive to the precise definition of the interface since this is "merely" the consequence of the 2D Fourier transform of the intermolecular potential. In essence, if one accepts that long-ranged dispersion interactions lead to 
an $\mathcal{O}(Q^3)$
correction in the denominator of the bulk structure factor then one must also accept that there is an $\mathcal{O}(q^2\ln q)$ correction to an {\it effective surface tension} $\sigma_\textup{eff}(q)$ defined according to (\ref{81}). Indeed, a straightforward calculation using (\ref{Gex}) and the definition (\ref{81}) yields a term $(\pi\epsilon a_\textup{LJ}^6m_0^2/8)\,q^2\ln(q\,a_\textup{LJ}\!)$ in $\sigma_\textup{eff}(q)$. The coefficient of the non-analytic $q^2\ln q$ term is precisely the same as that given in Refs.~\cite{Napio93,Mecke99,Blokhuis2009}; recall that $2m_0=\rho_l-\rho_v$ for a fluid. It is also instructive to return to the earlier Wertheim treatment of $G(z,z';q)$ \cite{Wertheim76}. Following the derivation in Appendix 4 of Ref.~\cite{Evans1979}, one finds that the 'one eigenvalue' ansatz of Wertheim yields a $\sigma_\textup{eff}(q)$ with exactly the same coefficient of the $q^2\ln q$ term as given above. One concludes that the Mecke-Dietrich analysis of the total structure factor $S^{tot}(
q)$ \cite{Mecke99} is equivalent to employing the Wertheim 'one eigenvalue' ansatz. However, we suspect that the same problems we exposed in our present treatment of short-ranged forces, regarding the definition, interpretation and measurement of an effective $\sigma(q)$, also arise for the case of long-ranged forces at higher wave-vectors.\\  

It would be instructive to go beyond the present DP approximation and consider the structure of pair correlations, the structure factor and the derivation of an interfacial model in the standard "$m^4"$ theory described by the potential $\phi(m)=-tm^2+um^4$, where $t\propto T_c-T$ is the deviation from the bulk critical temperature. In this case, the wave-vector dependence of both $G(z,z';q)$ and $S(z;q)$ can be determined. Many years ago, Zittartz provided an expression for $G({\bf r},{\bf r}')$ in MF \cite{Zittartz1967}. However, he did not show how this could be divided into bulk and excess contributions. Using methods different from Zittartz, we find that the local structure factor can be written as
\begin{equation}
S(z;q)=S_b(q)+\frac{2m_0 m'(z)}{\sigma q^2}\frac{S_b(q)}{S_b(0)}
\label{m4S}
\end{equation}
and appears to separate into bulk and excess contributions similar to the much simpler DP model. This result already tells us that the effective tension identified from the total structure factor, in the manner of (\ref{81}), $\sigma_\textup{eff}(q)=\sigma(1+q^2\xi_b^2)$, is similar to that obtained using the DP approximation and {\it increases} with $q$ - see (\ref{82}) and (\ref{X}). It may also be instructive to follow \cite{Parry2006a,Parry2007} and regard the DP model as the zeroth term in a perturbative description of the LGW model. These topics will be discussed in future work.\\

\acknowledgments

AOP acknowledges the support of the bank of Santander and UCIIIM for a Chair of Excellence and the EPRSC, UK, for grant "Creating macroscale 
effective interfaces encapsulating microstructural physics" EP/J009636/1. CR acknowledges support from grants FIS2010-22047-C05 (Ministerio de Educaci\'on y Ciencia) and MODELICO. RE acknowledges support from the Leverhulme Trust under award EM/2011-080.

\bibliography{wetting}

\end{document}